\documentclass[preprint,12pt]{revtex4}
\usepackage{graphicx,graphics,color,epsfig} 
\usepackage{physics}

\newcommand{\jb}{JB\ } 
\newcommand{\jbn}{JB} 

\newcommand{\ep}{\epsilon}

\begin{document}


\title{Strong-disorder renormalization group approach to the Anderson model using Raleigh-Schr\"odinger perturbation theory}

\author{Rachel Wortis}
\author{Eamonn Campbell}
\author{Donovan Allum}
\affiliation{Department of Physics \& Astronomy, Trent University, Peterborough, Ontario \ K9L0G2 \ Canada}

\date{\today}

\begin{abstract}
Previous work proposed a strong-disorder renormalization approach for the Anderson model, using it to calculate the density of states and the inverse participation ratio [Johri \& Bhatt, Phys.\ Rev.\ B {\bf 90} 060205(R) (2014)].  This is interesting because of the potential for expansion to higher dimensions and to interacting systems.  The original proposal used a non-standard perturbation theory approach which avoided degeneracies.  We implemented the same structure but with standard Rayleigh-Schr\"odinger perturbation theory.  Here degeneracies do arise, and we consider two approaches, one in which renormalization is suppressed if degeneracy is present and a second in which the most common form of degeneracy is handled using standard degenerate perturbation theory.   The version in which degeneracies are not handled performs similarly to the original proposal, and the addition of degeneracies provides some improvement.  
\end{abstract}








\maketitle

\section{Introduction}
\label{intro}

Disorder is present in many materials of current interest, either intrinsically or due to doping.
For example, most high temperature superconductors are achieved by doping insulating parent compounds, introducing charge carriers but also non-uniformity in the material parameters.
Disorder in non-interacting systems can give rise to localization effects\cite{Anderson1958,Abrahams1979}, and there is currently strong interest in the persistence of localization in interacting systems\cite{Gopalakrishnan2020,Abanin2019,Nandkishore2015}.
However, from a theory perspective, introducing disorder into techniques designed for studying strongly correlated electrons is often cumbersome.
Meanwhile, while real-space renormalization is of limited use in clean systems, strong disorder can generate a small parameter and hence a rational for renormalization.
Strong-disorder renormalization has a long history rooted in one-dimensional spin systems\cite{Ma1979,Dasgupta1980,Fisher1994}, and has since expanded to encompass a broad spectrum of approaches addressing a wide diversity of systems\cite{Igloi2005,Igloi2018}.

Many renormalization procedures, including the original Wilson approach\cite{Wilson1974}, are designed around energy scales,
and strong-disorder renormalization as applied to the Anderson model has often focused on the ground state.\cite{Monthus2009,Mard2014,Mard2017}
However, localization is a property of wavefunctions which is not monotonically related to energy.
A particular large-disorder renormalization group approach to the Anderson model was proposed in Ref.\ \cite{Johri2014}, 
hereafter referred to as \jbn.
This approach is organized around the level of localization of the single-particle states removed, working sequentially from the most local to the least.
The authors demonstrated that the method flows toward strong disorder and benchmarked the density of states and inverse participation ratio against exact results with encouraging results.
The method used perturbation based on the small parameter of hopping amplitude divided by difference in site potentials for neighboring sites.
However, the specific perturbation scheme used diverged from standard perturbation theory.

Motivated by the potential to adapt this approach to higher dimensions and interacting systems, 
we examine here whether the use of standard Rayleigh-Schr\"odinger perturbation improves the performance of the \jb approach.
Notably, the non-standard approach taken in \jb removes the possibility of degeneracies.  While this simplifies the calculation, it limits the capacity of the procedure to capture more extended states.
We demonstrate that consistent use of standard perturbation theory results in improvement in the match of the inverse participation ratio to exact results.

Section \ref{method} describes the method, including reviewing the original proposal and specifying three versions we have implemented.
Section \ref{results} presents our results and discusses them relative to several points of comparison.

\section{Method}
\label{method}

This section begins with a description of the model studied and the quantities calculated.  
It then presents the renormalization procedure, reviewing the approach taken in \jb and highlighting where it diverges from standard perturbation theory.  Two alternative solution methods used as points of comparison are also described.

The Anderson model is a tight-binding model for non-interacting electrons with diagonal disorder.
\begin{eqnarray}
H &=& \sum_i E_i c_i^{\dag} c_i + \sum_i t_{i,i+1} (c_i^{\dag} c_{i+1} + c_{i+1}^{\dag} c_i)
\end{eqnarray}
where $c_i^{\dag}$ and $c_i$ are the creation and annihilation operators for site $i$, $t_{i,i+1}$ is the hopping amplitude between nearest neighbor sites (initially $t_{i,i+1}=-1\ \forall \ i$), and the site potentials $E_i$ are chosen from a flat distribution of width $W$.
We focus here on one-dimensional systems with periodic boundary conditions.

The main quantities calculated are the density of states (DOS) and the inverse participation ratio (IPR).
The DOS is the probability distribution of the single-particle energies, and the IPR is a measure of the localization of the single-particle states.  
Each single-particle state may be expressed in the basis of Wannier states $\ket{i}\equiv c_i^{\dag}\ket{vacuum}$.
For a particular state $\ket{\psi_{\alpha}}=\sum_i w_{\alpha,i}\ket{i}$, the IPR
\begin{eqnarray}
I_{\alpha} &=& \frac{\sum_i |w_{\alpha,i}|^4}{\left( \sum_i |w_{\alpha,i}|^2 \right)^2}.
\end{eqnarray}
The IPR as a function of energy for an ensemble of systems is generated by assigning $I_{\alpha}$ of each state to a bin according to the energy of the state $E_{\alpha}$, and then averaging the values in each bin. 

The renormalization procedure begins by evaluating, for each pair of adjacent sites, the ratio of hopping amplitude to the difference in site potentials, $m_{i,i+1}\equiv t_{i,i+1}/|E_i-E_{i+1}|$, referred to as the bond strength.  
These bonds are then sorted into weak and strong, according to whether they are above or below a chosen cutoff, $m_0$.  
The system is then divided into clusters, where a cluster is defined as a collection of adjacent sites all connected by strong bonds and linked to the rest of the system at each end by a weak bond.

Regarding the choice of the bond cutoff, $m_0$, a basic starting point is that it should be less than one.  
Beyond this, smaller values generate larger clusters and correspondingly more accurate results,
while larger values generate smaller clusters and hence faster results.

As a point of comparison for the renormalization results, we also consider an approach which stops here:  The weak bonds are set to zero resulting in an ensemble of independent clusters, and each cluster is then solved exactly.

In the renormalization, the next step is to identify the single-site cluster which is least strongly coupled to its neighbors, specifically that for which the connectivity $m_{i,i+1}^2+m_{i-1,i}^2$ is minimum. 
The state $\ket{C}$ on the identified site and its associated energy $E_C$ are renormalized by treating the couplings $t_{L,C}$ and $t_{C,R}$ to the left and right neighbor sites $|L\rangle$ and $|R\rangle$ as a perturbation using standard time-independent perturbation theory to first order in in $t/\Delta E$.  
\begin{eqnarray} 
|C'\rangle &=& |C\rangle - {t_{L,C} \over E_L-E_C} |L\rangle - {t_{C,R} \over E_R-E_C}  \ket{R}
\label{1st-single-site-eqn}\\
E_C' &=& E_C - {t_{L,C}^2 \over E_L-E_C} 
		- {t_{C,R}^2 \over E_R-E_C}
\end{eqnarray}
The energy $E_C'$ and the IPR $I$ of $\ket{C'}$ are recorded, and the site is removed from the system.
The renormalized system includes the left and right neighbor sites similarly perturbed by their coupling to the removed site and an effective coupling between them.
\begin{eqnarray}
|L'\rangle &=& |L\rangle - {t_{L,C} \over E_C-E_L} \ket{C}\\		
E_L' &=& E_L - {t_{L,C}^2 \over E_C-E_L} \\		
|R'\rangle &=& |R\rangle - {t_{C,R} \over E_C-E_R} \ket{C} \\		
E_R' &=& E_R - {t_{C,R}^2 \over E_C-E_R} \\
t_{LR} &=& \langle L'|H|R'\rangle \nonumber \\
&=& -t_{L,C} t_{C,R} \left( {1 \over E_C-E_L} + {1 \over E_C-E_R} \right) 
\label{last-single-site-eqn}
\end{eqnarray}
This process continues until all one-site clusters have been removed.
Up to this point, all versions we have implemented are identical and the same as that in \jbn.

\begin{figure}[htbp] 
\includegraphics[width=2 in]{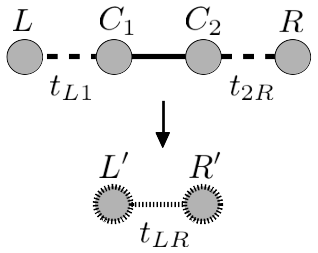} 
\caption{\label{diagram}Schematic diagram of the procedure for the case of a two-site cluster.  The cluster sites $C_1$ and $C_2$ are connected by a strong bond, with weak bonds to the left $L$ and right $R$ sites respectively.  The left and right states are renormalized, becoming $L'$ and $R'$ now connected by a new effective coupling $t_{LR}$.}
\end{figure}

The next step is to proceed to clusters of two sites and then three sites, etc.
Fig.\ (\ref{diagram}) shows a schematic diagram of the process for the case of a two-site cluster.
For a cluster of $n$ sites ($C_1$ to $C_n$), the exact eigenstates $|\psi_q\rangle=\sum_{j=1}^{n} c_{q,j} |C_j\rangle$ are determined and then, as for the single-site case, these are renormalized by treating the hopping to the left and right neighbor sites as a perturbation.
It is at this point that the Rayleigh-Schr\"odinger approach diverges from \jbn.
Table \ref{table1} lists the equations for renormalization of the cluster states and energies as well as the neighbor states and energies.
The center column has the equations used by \jb and the right column shows the equations obtained from a standard Rayleigh-Schr\"odinger calculation.
We implement all renormalization approaches to linear order in $t/\Delta E$, and at this order the basis remains orthogonal (see \ref{app:basis}).
Weights on the original basis are used when calculating IPR values.

\begin{table*}
\begin{eqnarray}
\begin{array}{l |l |l}
& {\rm JB \ RG \ equations} & {\rm RS \ RG \ equations} \\ \hline
|\psi_q'\rangle = |\psi_q \rangle + ...
& 
- {t_{L,1} \over E_L-E_1} c_{q,1} |L\rangle 
- {t_{n,R} \over E_R-E_n} c_{q,n} |R\rangle 
& 
- {t_{L,1} \over E_L-E_q} c_{q,1} |L\rangle 
- {t_{n,R} \over E_R-E_q} c_{q,n} |R\rangle
\\
E_q' = E_q + ...
& -{t_{L,1}^2 \over E_L-E_1} c_{q,1} 
		- {t_{n,R}^2 \over E_R-E_n} c_{q,n}
& - {t_{L,1}^2 \over E_L-E_q} c_{q,1}^2 
		- {t_{n,R}^2 \over E_R- E_q} c_{q,n}^2 
\\
|L'\rangle = |L\rangle + ...
& - {t_{L,1} \over E_1-E_L} \left( \sum_{q=1}^n c_{q,1} \right) |C_1\rangle
& - \sum_{q=1}^n {t_{L,1} \over E_q-E_L} c_{q,1} |\psi_q\rangle 
\\
E_L' = E_L + ...
& - {t_{L,1}^2 \over E_1-E_L} \left( \sum_{q=1}^n c_{q,1} \right)
& - \sum_{q=1}^n {t_{L,1}^2 \over E_q-E_L} c_{q,1}^2 
\\
|R'\rangle = |R\rangle +...
& - {t_{n,R} \over E_n - E_R} \left( \sum_{q=1}^n c_{q,n}\right) |C_n\rangle
& - \sum_{q=1}^n {t_{n,R} \over E_q-E_R} c_{q,n} |\psi_q \rangle 
\\
E_R' = E_R + ...
& - {t_{n,R}^2 \over E_n-E_R} \left( \sum_{q=1}^n c_{q,n}\right)
& - \sum_{q=1}^n {t_{n,R}^2 \over E_q - E_R} c_{q,n}^2
\\
t_{LR} = 
& - t_{L1} t_{nR} \left( 
	\frac{\sum_{q=1}^n c_{q,1}}{E_1-E_L}
	+ \frac{\sum_{q=1}^n c_{q,n}}{E_n-E_R} \right){\ddag}
& - \sum_{q=1}^n t_{L,1} t_{n,R} c_{q,1} c_{q,n} 
	\left( {1 \over E_q - E_L} + {1 \over E_q-E_R} \right)
\nonumber
\end{array}
\end{eqnarray}
\caption{\label{table1}Equations for the renormalized states and energies of the cluster and of its neighbors, as well as the hopping amplitude between the left and right sites in the renormalized system.
Sites $L$ and $R$ are to the left and right, respectively, of cluster sites $C_j$, $j=1,n$.
$t_{L,1}$, and $t_{n,R}$ are the hopping amplitudes between $L$ and $C_1$ and between $C_n$ and $R$. 
$E_L$, $E_R$, $E_1$ and $E_n$ are the (pre-renormalization) energies of the sites $L$, $R$, $C_1$, and $C_n$, respectively.
$E_q$ is the (pre-renormalization) energy of the cluster eigenstate $|\psi_q\rangle=\sum_{j=1}^{n} c_{q,j} |C_j\rangle$.
$\ddag$As discussed in the text, \jb does not provide an expression for $t_{LR}$ for $n>1$.  We use this expression in our implementation of their work.
}
\end{table*}

Both sets of equations in Table \ref{table1} reduce to Eqns.\ (\ref{1st-single-site-eqn})-(\ref{last-single-site-eqn}) in the case $n=1$.
Nonetheless, the two sets of equations differ significantly.  Two differences in particular stand out.
First, the denominators:  In both cases, the denominators contain the difference between the energy of a neighboring site ($E_L$ or $E_R$) and an energy associated with the cluster.  In the Rayleigh-Schr\"odinger case, this second energy is a cluster-state energy $E_q$, whereas in the \jb case this second energy is the energy of the adjacent end-site in the cluster ($E_1$ or $E_n$).
Second, the renormalization of the neighbor states $\ket{L}$ and $\ket{R}$ in the Rayleigh-Schr\"odinger case involves the cluster state $\ket{\psi_q}$ (and hence all sites in the cluster), whereas in the \jb case only the sites $\ket{C_1}$ and $\ket{C_n}$ at the two ends of the cluster are involved.

It might seem surprising not to use standard Rayleigh-Schr\"odinger perturbation theory, but there is a practical advantage to the \jb approach.   
Namely, because the denominators always involve precisely the bond strengths used to identify the cluster, there is never a case of degeneracy.
Meanwhile, in the Rayleigh-Schr\"odinger case, it is possible for a cluster state energy $E_q$ to be close to the energy of one of the neighboring sites, such that the terms shown in Table \ref{table1} diverge and a degenerate perturbation calculation is required.

While the lack of degeneracy makes the \jb approach simpler, it also limits its ability to capture more extended states.
With the intent to understand how much difference this makes, we implement two versions of the  Rayleigh-Schr\"odinger calculation.  
In one, SDRG-RSnd, only clusters with no degeneracy are renormalized.
Any cluster in which degeneracy arises is not renormalized but instead handled as an independent cluster (as if the bonds at the two ends of the cluster had zero strength).
In the second, SDRG-RSdg, renormalization is carried out, using standard degenerate perturbation theory, in clusters where a single cluster state is degenerate with just one of the two neighbor sites (see \ref{app:deg}).
In more complicated cases, such as degeneracy of a cluster state with both neighbor sites or degeneracy of more than one state of a given cluster, still no renormalization is done.
The simple case accounts for the bulk of the degeneracies, and the more complicated cases are very rare, as quantified below in Section \ref{results}.

The different approaches to the renormalization of states and energies also result in distinct expressions for the  hopping $t_{LR}$ between the left and right sights in the renormalized system.  
\jb defines 
\begin{eqnarray}
t_{LR} \equiv \bra{L'}H\ket{R'} \label{tLRdefn}
\end{eqnarray}
but provides no more explicit expression for it in either single-site or multi-site clusters.
In the case of a single-site cluster, both $\ket{L'}$ and $\ket{R'}$ have nonzero weight on the original site $\ket{C}$, resulting in a nonzero value of $t_{LR}$ as shown in Eqn.\ (\ref{last-single-site-eqn}).  
Moreover, in the case of a multi-site cluster in the Rayleigh-Schr\"odinger version, both $\ket{L'}$ and $\ket{R'}$ have nonzero weight on all sites in the cluster, again giving rise to a nonzero value of $t_{LR}$.
However, in the \jb approach, $\ket{L'}$ has weight only on $\ket{C_1}$, and $\ket{R'}$ has weight only on $\ket{C_n}$, such that when $n>1$, $t_{LR}=0$.  
Meanwhile, \jb Fig.\ 3 shows the generation of nonzero bonds for all cluster sizes, indicating that 
\jb used something different from Eqn.\ (\ref{tLRdefn}) for $t_{LR}$ in the case of multi-site clusters.  
The expression for $t_{LR}$ in the \jb column of Table \ref{table1} is something we have constructed following the form of the other \jb equations which reduces to the one-site cluster result in the limit $n\to 1$ and produces results similar to those shown in \jb as discussed in Section \ref{results} below.

In summary, we present the results of three renormalization approaches:  
SDRG-JB is our implementation of the method described in \jb to first order in $t/\Delta E$.  
SDRG-RSnd uses standard Rayleigh-Schrodinger perturbation theory to first order in $t/\Delta E$, but for clusters in which degeneracy arises no renormalization is performed.  
Finally, SDRG-RSdg uses standard Rayleigh-Schrodinger perturbation theory to first order in $t/\Delta E$ and handles the most common form of degeneracy.
We consider two choices of bond cutoff.

We compare the DOS and IPR results of these renormalization approaches to two alternatives:  an exact solution and the ensemble of independent-clusters described above.  
While the exact solution provides the ultimate benchmark, the independent-cluster solution gives a sense for the impact of the renormalization.

As diagnostics of the renormalization procedures, and for comparison with \jbn, we also calculate the fraction of the sites in the system which are removed at a given cluster size $N_{cl}$, the distribution of bond strengths after all clusters of a given size have been removed, and the number of degeneracies encountered.

\section{Results and Discussion}
\label{results}


\begin{figure}[htbp] 
\includegraphics[width=\columnwidth]{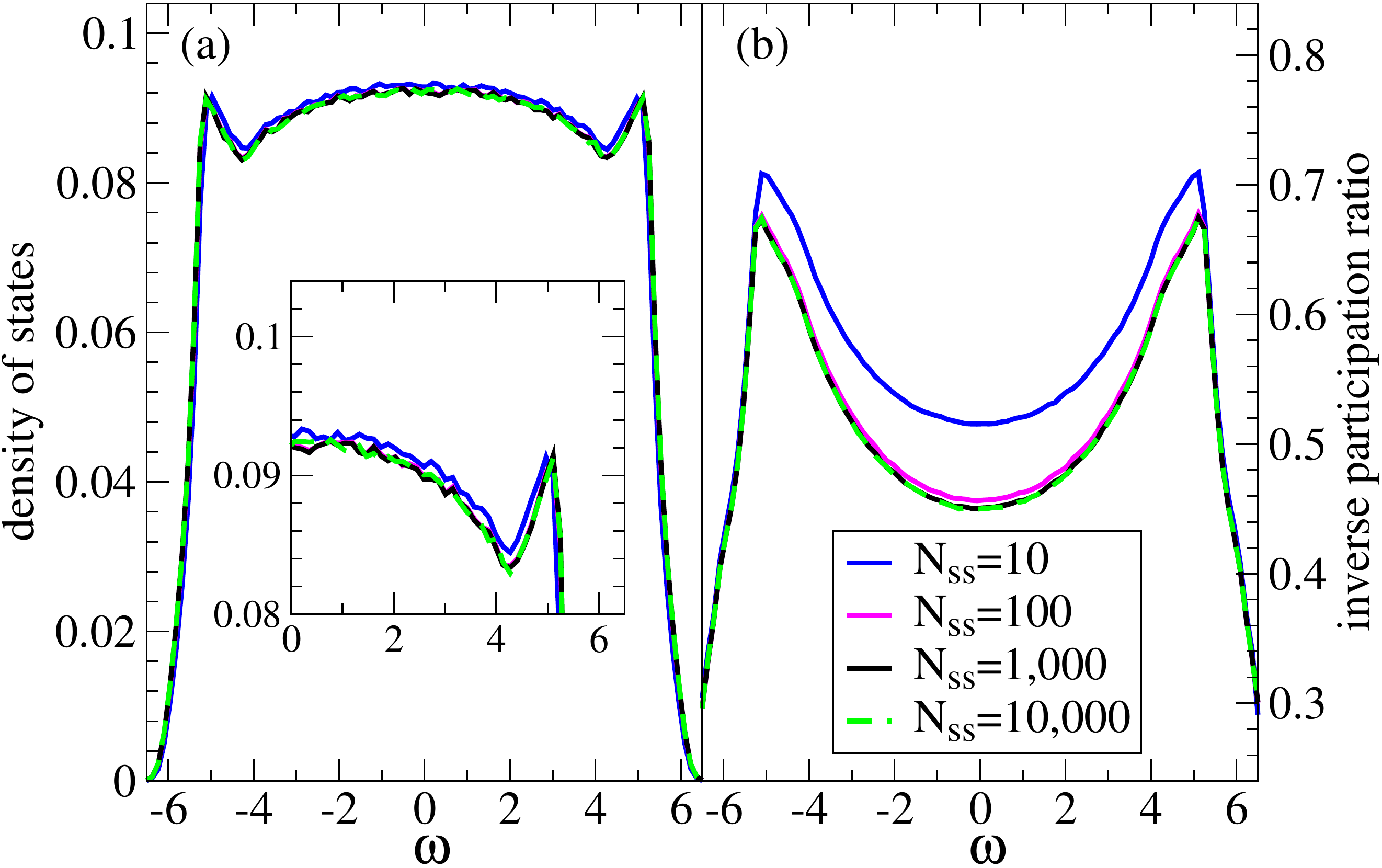} 
\caption{\label{dosipr_exact}Exact density of states and inverse participation ratio at disorder strength $W=10$.  Results for a range of system sizes, with the product of the system size $N_{ss}$ and the number of disorder configurations $N_{dc}$ held constant at $10^7$.}
\end{figure}

This section begins with a brief examination of the exact results and then presents the independent-cluster and renormalization results for two values of the bond cutoff.

Fig.\ \ref{dosipr_exact} shows the exact DOS and IPR results for different system sizes, $N_{ss}$.
The number of disorder configurations, $N_{dc}$, is chosen such that the total number of sites is the same for all cases.
The key message here is that for disorder strength $W=10$ (the value used in \jbn), the results are indistinguishable for $N_{ss}>10^3$.

\begin{figure}[htbp] 
\includegraphics[width=\columnwidth]{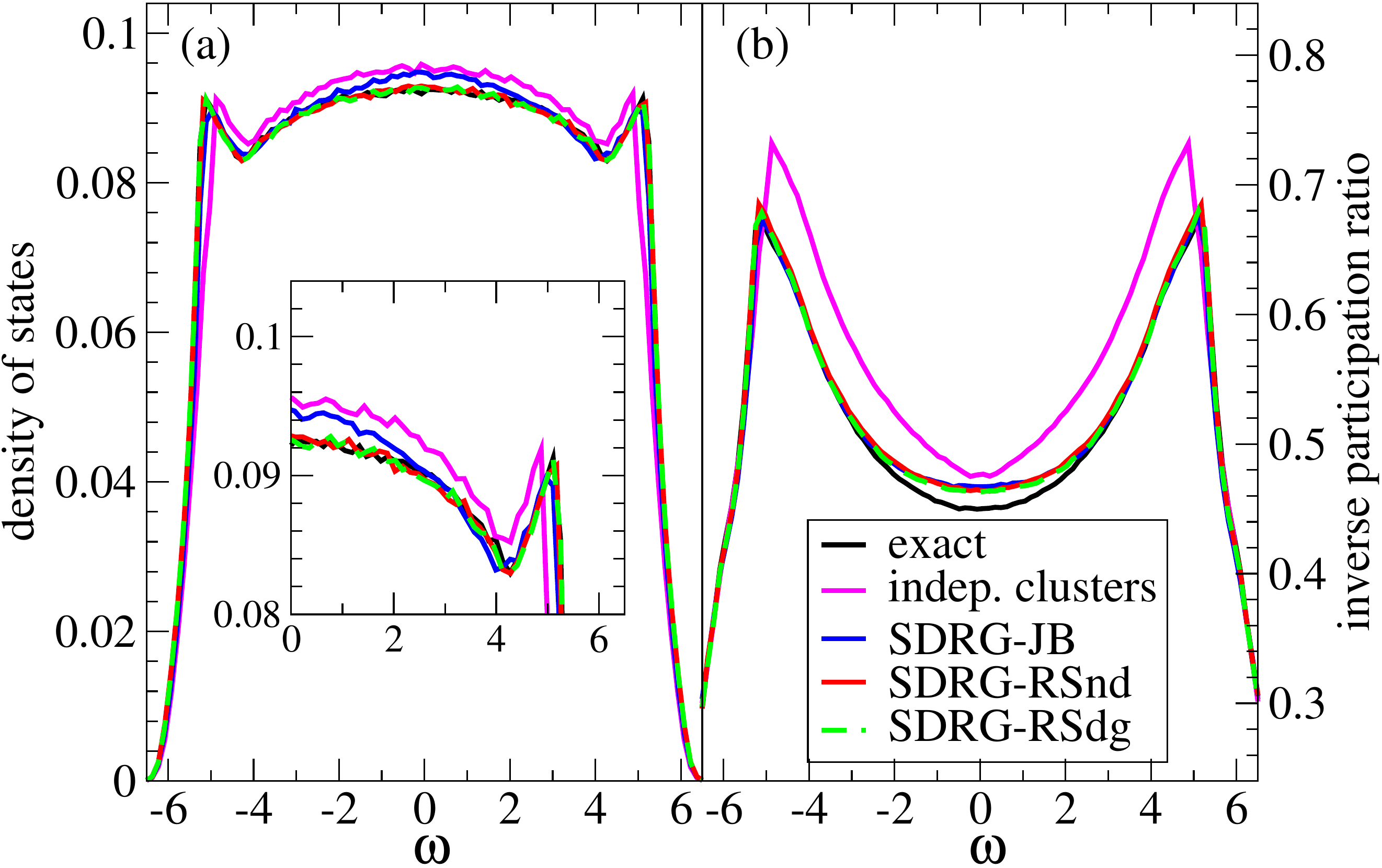} 
\caption{\label{dosipr_bc02}Density of states and inverse participation ratio results at disorder strength $W=10$ using bond cutoff $m_0=0.2$.  $N_{ss}=10^4$ and $N_{dc}=10^3$.}
\end{figure}

Fig.\ \ref{dosipr_bc02} shows the DOS and IPR results for disorder strength $W=10$ from the exact, independent-cluster, and renormalization approaches using bond cutoff $m_0=0.2$, system size $10^4$, and 1000 disorder configurations.
The independent-cluster results are qualitatively similar to the exact results but clearly distinct quantitatively, both in magnitude and in the location of the peaks.\cite{Johri2012a,Johri2012b}
For the DOS, the SDRG-RSdg results are essentially the same as those of SDRG-RSnd, and both are very close to the exact, while the SDRG-JB results diverge, especially near the band center.
For the IPR, all the renormalization methods give similar results, following the exact results closely at the band edges but deviating in the band center.

\begin{figure}[htbp] 
\includegraphics[width=\columnwidth]{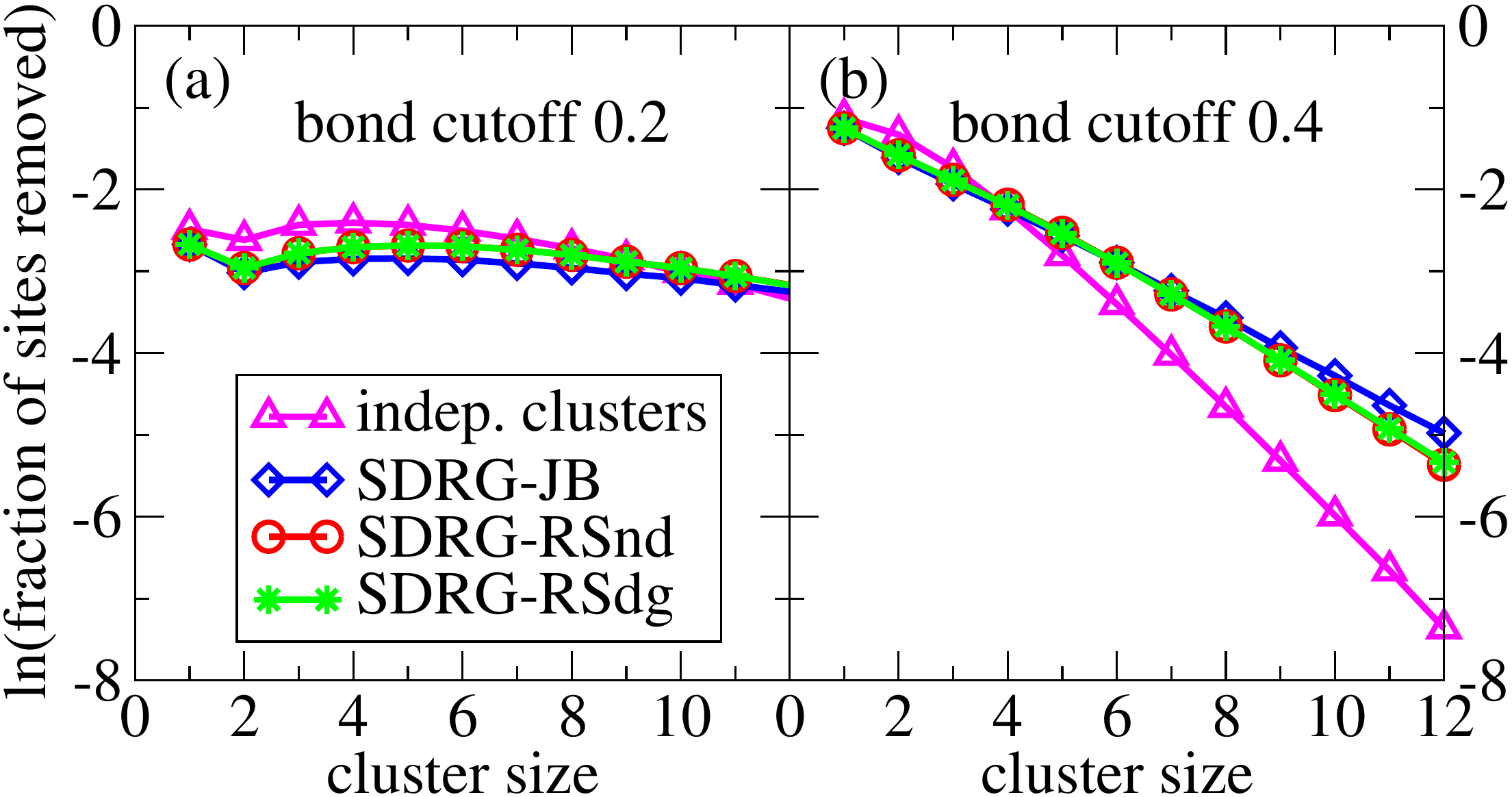} 
\caption{\label{sts}The fraction of sites removed at each cluster size.  Disorder strength $W=10$, $N_{ss}=10^4$, and $N_{dc}=10^3$.}
\end{figure}

Fig.\ \ref{sts} shows the fraction of sites removed at each cluster size for the independent-cluster approach, SDRG-JB, and both the SDRG-RS approaches.  
In general, one might expect that this fraction is small at small cluster size (because, although there are many such clusters, each contains only a small number of sites) and small at large cluster size (because, although each contains many sites, there are only a few such clusters) with a peak in between.  
The structure is indeed seen Fig.\ \ref{sts}(a) with the notable exception of an enhanced value at cluster size one.  
This enhancement reflects the fact that, when the bond cutoff approaches $1/W$, bonds adjacent to a weak bond have an increased probability of also being weak:  The two sites surrounding a weak bond must be near opposite edges of the band.  If $m_{i,i+1}<m_0$, then $E_i$ and $E_{i+1}$ are near (opposite) band edges, and therefore $m_{i-1,i}$ and $m_{i+1,i+2}$ are more likely to also be weak.

For a given value of the bond cutoff, the result for the independent-cluster approach represents the maximum rate of removal at small cluster sizes.  
Any renormalization scheme producing renormalized bonds which are strong combines two smaller clusters into one larger one, removing weight from the small cluster region of this diagram and shifting it to higher cluster sizes.
This shift is indeed visible in all three renormalization schemes.
The results for the two SDRG-RS approaches are very similar, and the SDRG-JB approach shows slightly greater shift to large clusters (reduction in small clusters), corresponding to a higher rate of generation of strong bonds.

\begin{figure}[htbp] 
\includegraphics[width=\columnwidth]{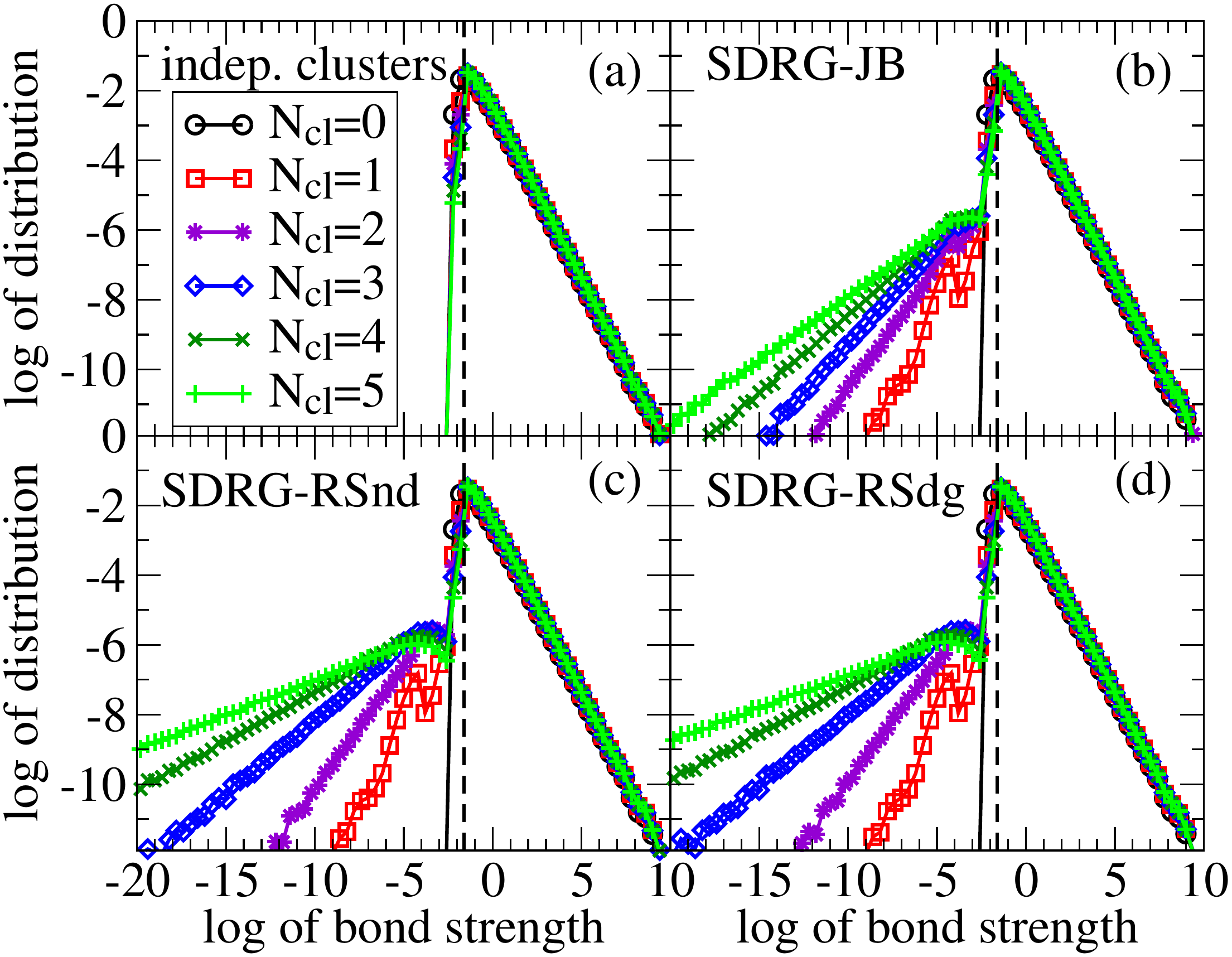} 
\caption{\label{dbs_bc02}Logarithm of the distribution of the log of the bond strengths at disorder strength $W=10$ using bond cutoff $m_0=0.2$.
$N_{ss}=10^4$ and $N_{dc}=10^3$.}
\end{figure}

Fig.\ \ref{dbs_bc02} shows the log of the distribution of bond strengths for each approach.  
In the independent-cluster approach, no new bonds are generated, so the distribution shows just a monotonic decline in the weight below the bond cutoff.
All the SDRG approaches do generate new bonds, most of which are weak, reflected in the evolution of their distributions toward greater weight at low values and hence a flow toward the strong-disorder limit.  
The rate at which weak bonds are generated is slower in SDRG-JB than in the two SDRG-RS approaches, consistent with the data on the fraction of sites removed (Fig.\ \ref{sts}).

For all quantities calculated with this bond cutoff, the difference between SDRG-RSnd and SDRG-RSdg is negligible:  handling cases of degeneracy makes very little difference.  
For this relatively high disorder strength ($W=10$) and low bond cutoff ($m_0=0.2$), degeneracy arises in less than 0.6\% of states, and multiple degeneracy arises in 0.002\% of cases.
Nonetheless, in Fig.\ \ref{dosipr_bc02}(b) the SDRG-RSdg curve is very slightly lower (and closer to the exact) at the band center, reflecting the particular significance of degeneracy in the formation of more extended states.


\begin{figure}[htbp] 
\includegraphics[width=\columnwidth]{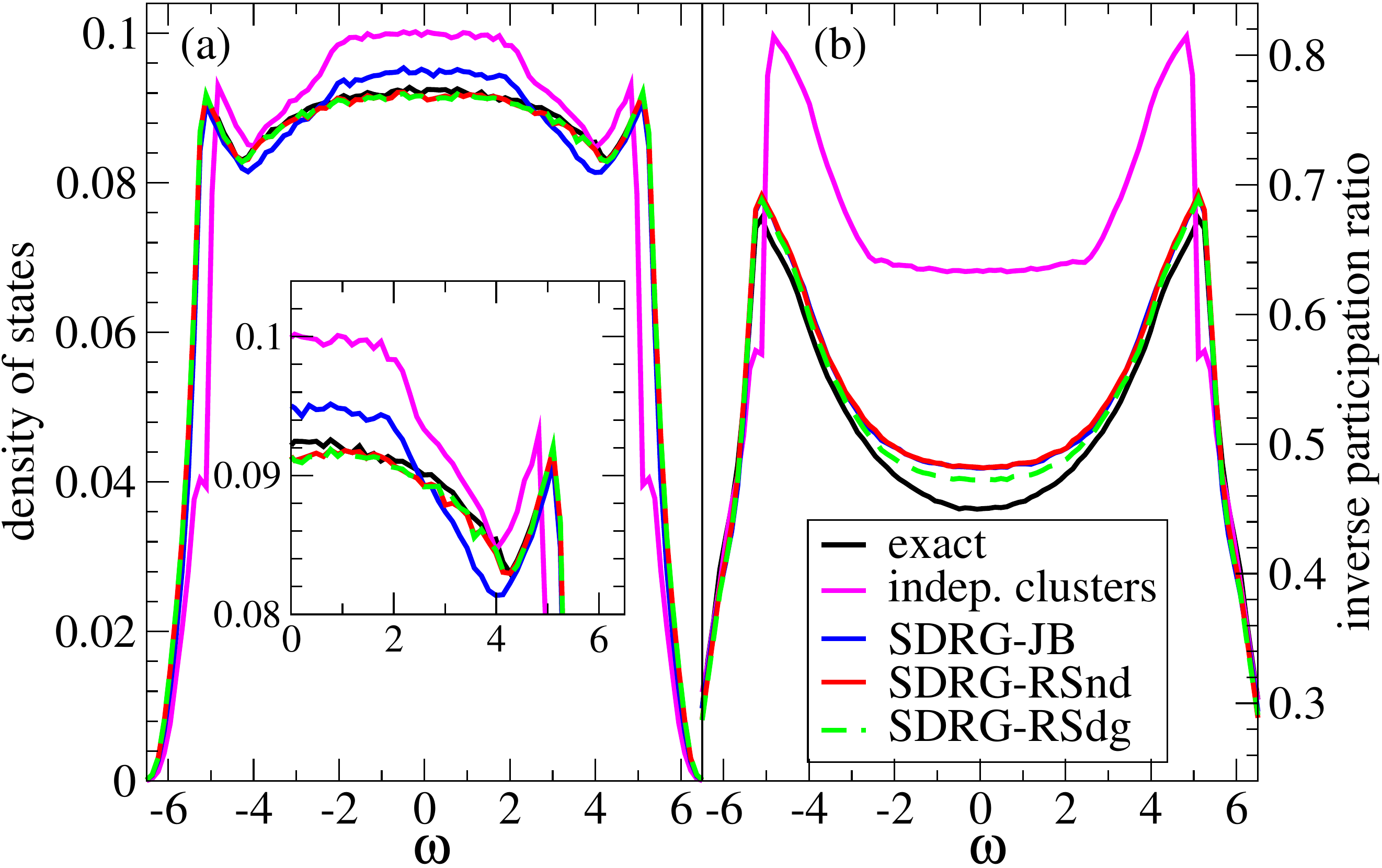} 
\caption{\label{dosipr_bc04}Density of states and inverse participation ratio results at disorder strength $W=10$ using bond cutoff $m_0=0.4$.  $N_{ss}=10^4$ and $N_{dc}=10^3$.}
\end{figure}

\begin{figure}[htbp] 
\includegraphics[width=\columnwidth]{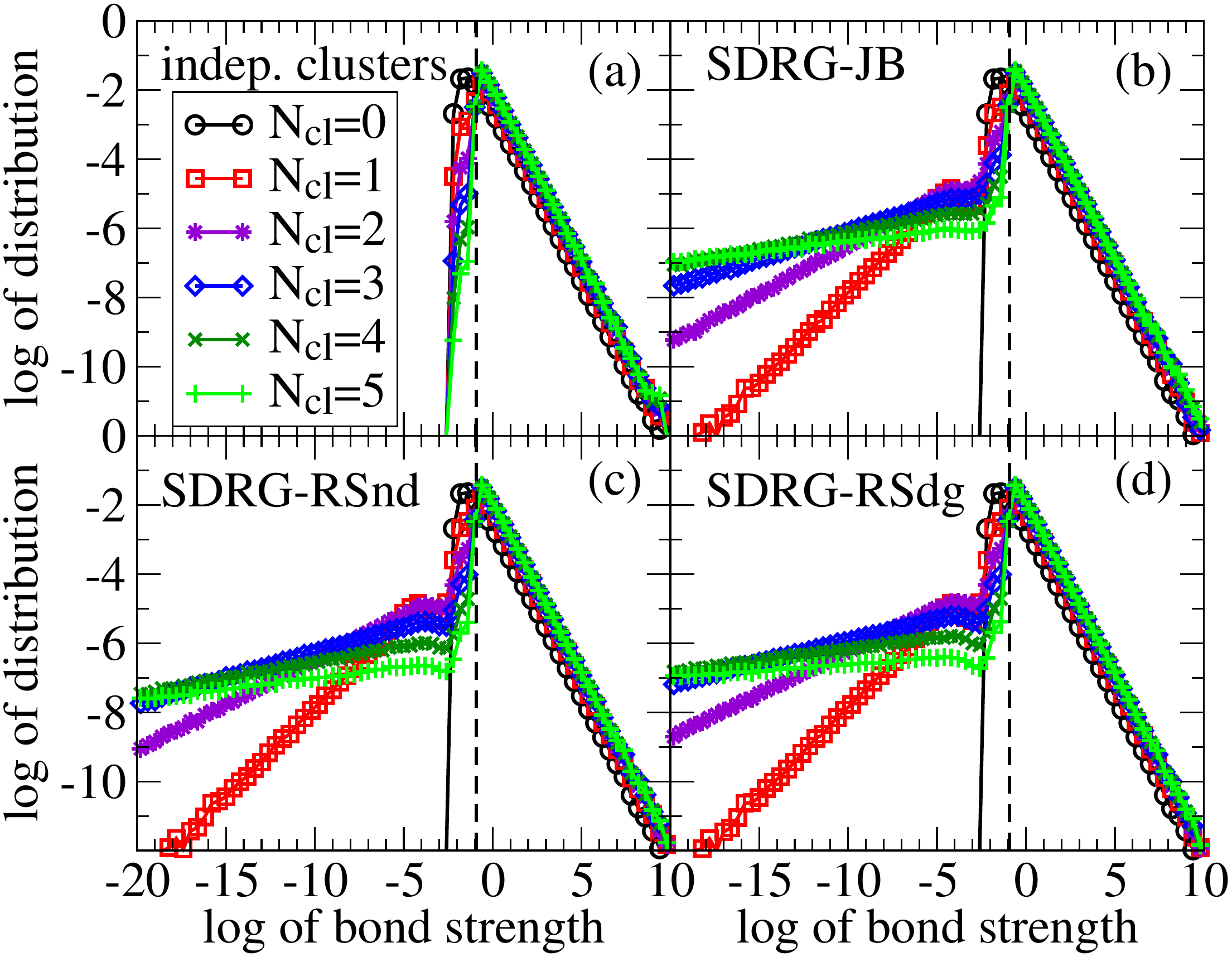} 
\caption{\label{dbs_bc04}Logarithm of the distribution of the log of the bond strengths at disorder strength $W=10$ using bond cutoff $m_0=0.4$.
$N_{ss}=10^4$ and $N_{dc}=10^3$.}
\end{figure}

A motivation for studying this method is the potential to extend it to interacting systems.
Given that computational time grows exponentially with size in interacting systems,
the performance of the method at larger bond cutoff with correspondingly smaller cluster sizes is of interest.

Figures \ref{dosipr_bc04}, \ref{sts}(b), and \ref{dbs_bc04} show the corresponding results for bond cutoff, $m_0=0.4$.
In Fig.\ \ref{dosipr_bc04}(a), the independent-cluster approach results are significantly farther from exact than for $m_0=0.2$.
The SDRG-JB results are also worse while both SDRG-RS results remain quite close to the exact.
For the IPR, Fig.\ \ref{dosipr_bc04}(b), the independent-cluster results are very far from the exact, and, unlike with $m_0=0.2$, all the SDRG results are higher than the exact results not just at the band center but all the way out to the peaks.
The SDRG-JB and SDRG-RSnd results are again similar to each other, while the difference between these and the SDRG-RSdg results is larger than for $m_0=0.2$.  
The impact of handling degeneracies is more significant because there are more cases of degeneracy.  
Nonetheless, degeneracy still only arises in less than 2\% of states, and multiple degeneracy arises in 0.02\% of cases.

In \ref{sts}(b), the fraction of sites removed shows a monotonic decrease.
For this larger value of the bond cutoff, the fraction of bonds which are weak is much larger, resulting in a correspondingly greater proportion of small clusters.
The renormalization approaches still generate strong bonds, resulting in a shift of weight to larger cluster sizes.
In Fig.\ \ref{dbs_bc04}, the distribution of bond strengths shows more rapid generation of weak bonds than for $m_0=0.2$.
In both the fraction of sites removed and the distribution of bond strengths, the results for the two SDRG-RS approaches are similar, and SDRG-JB produces more strong bonds resulting in greater shift toward larger cluster sizes in Fig.\ \ref{sts}(b) than (a) and less rapid generation of weak bonds in Fig.\ \ref{dbs_bc04} than \ref{dbs_bc02}.

We now compare the results of our SDRG-JB approach with the results shown in \jbn.
Comparing Fig.\ \ref{sts}(a) here with \jb Fig.\ 1, an inconsistency is immediately apparent:  The data labeled ``$w=10$, $m_0=0.2$'' in \jb show a monotonic decline unlike our Fig.\ \ref{sts}(a).  Particularly puzzling, the fraction of sites removed at small cluster sizes in \jb is much larger than even the independent-cluster approach results in our Fig.\ \ref{sts}(a).  
Similarly, the rate of generation of weak bonds in \jb Fig.\ 3 is much more rapid than in our Fig.\ \ref{dbs_bc02}.
It was partly in response to this discrepancy that we considered the case of bond cutoff $m_0=0.4$.
Indeed, the SDRG-JB results in our Fig.\ \ref{sts}(b) bear a strong resemblance to the curve labeled ``$w=10$, $m_0=0.2$'' in \jb Fig.\ 1, and correspondingly the SDRG-JB results in our Fig.\ \ref{dbs_bc04} show a strong resemblance to \jb Fig.\ 3.  
This suggests there may be a factor of two difference between their implementation of the bond cutoff and ours.  
Consistent with this, our IPR results for bond cutoff $m_0=0.4$ (Fig.\ \ref{dosipr_bc04}(b)) are similar to those labeled $m_0=0.2$ in \jb Fig.\ 2 (see \ref{app:compare}). In particular the vertical offset, which is maximum at the band center, extends all the way to the peaks near the band edge.
This is in contrast to our IPR results for bond cutoff $m_0=0.2$ (Fig.\ \ref{dosipr_bc04}(a)) in which the difference from the exact results is only readily apparent near the band center.

However, a factor of two difference in the bond cutoff does not fully resolve the differences between our results and those in \jbn.  
In particular, the DOS results we find with the SDRG-JB approach for bond cutoff $m_0=0.4$ deviate strongly from the exact results, unlike the DOS results shown in \jb Fig.\ 2 (see \ref{app:compare}).
Further, the results in \jb labeled $m_0=0.05$ are puzzling:  All initial bonds for disorder strength $W=10$ would be greater than this cutoff, such that the whole system should be a single cluster, and the results should be identical to the exact results.  This is true even if the cutoff is doubled.
It is unfortunate that the \jb paper does not specify (i) the form of the renormalized hopping amplitude $t_{LR}$ in the case of multi-site clusters, (ii) the approach they took to the loss of orthogonality in the basis, or (iii) why they chose to address this given that it only arises beyond first order in $t/\Delta E$.

Nonetheless, setting aside the relatively minor quantitative differences between our results and those of \jb as well as between the different approaches we've implemented, an important question is whether any of these provides a practical advantage.
For the one-dimensional non-interacting systems studied here, if the desire is to generate the DOS and IPR across the full spectrum maximizing accuracy and speed, these SDRG approaches appear to be less effective than simply applying exact methods to an ensemble of smaller systems.
In particular, comparing Fig.\ \ref{dosipr_bc02} and Fig.\ \ref{dosipr_bc04} with Fig.\ \ref{dosipr_exact}, we note that none of the RG approaches is as close to the converged exact result as the exact results for $N_{ss}=100$.
These exact results for 100-site systems take only a few minutes of computational time, roughly two orders of magnitude less than the exact results for systems of 10$^4$ sites, and also significantly less than the time to run the renormalization code on the larger systems.


\section{Conclusion}
\label{conclu}

In conclusion, we have implemented three versions of strong-disorder renormalization to calculate the DOS and IPR of the one-dimensional Anderson model and compared with exact results and with the results of a simple ensemble of independent clusters.
The first renormalization scheme SDRG-JB followed both the framework and the renormalization equations proposed in \jbn, keeping terms first order in $t/\Delta E$.  
The other two renormalization approaches employed the same framework but the renormalization equations used standard Rayleigh-Schrodinger perturbation theory to first order in $t/\Delta E$.
In contrast to the \jb scheme, this leads to the potential for degeneracies.
SDRG-RSnd detects cases of degeneracy and does not apply renormalization to those clusters, 
whereas SDRG-RSdg uses standard degenerate perturbation theory to handle most, although not all, cases of degeneracy.

Comparison of the independent-cluster approach and SDRG results in Fig.\ \ref{dosipr_bc02} and Fig.\ \ref{dosipr_bc04} demonstrates that the renormalization is providing significant improvement, especially for the IPR.
The SDRG results for both DOS and IPR come close to the exact results, especially near the edges of the band.

We find that properly handling degeneracies does improve the results of the renormalization, that is the SDRG-RSdg results are closer to the exact than those of SDRG-RSnd, particularly for the IPR.
However, this improvement is relatively minor 
and SDRG-RSnd does not show an improvement relative to the ad-hoc method of SDRG-JB.

In comparing our results with those presented in \jbn, we note the results labeled in \jb as $m_0=0.2$ are inconsistent with our results for bond cutoff $m_0=0.2$ but show similarities with our bond cutoff $m_0=0.4$ results.  

The renormalization schemes do all flow to strong disorder, and they all show a significant improvement relative to the simple ensemble of independent clusters, and they are particular effective in capturing the IPR at the band edges.
However, the high disorder values which justify the use of the strong-disorder renormalization also result in a small localization length, and we find that in non-interacting systems an exact treatment of large numbers of small systems is faster and closer to the converged exact results than any of the SDRG schemes.
Extension to interacting systems, in which exact solution of even small clusters is numerically demanding, may still hold potential.

\section*{Acknowledgments}
We acknowledge support by the National Science and Engineering Research Council (NSERC) of Canada.
This work was made possible by the facilities of the
Shared Hierarchical Academic Research Computing Network
(SHARCNET). 
RW acknowledges helpful conversations with Sonika Johri and Ravin Bhatt, 
and thanks Malcolm Kennett for feedback on the manuscript.

\appendix

\section{Degeneracy}
\label{app:deg}


\subsection{Defining degeneracy}

In our calculation, a cluster state $\ket{\psi_m}=\sum_{j=1}^n w_{m,j} \ket{Cj}$ with energy $E_m$ is determined to be degenerate with the right neighbor site when 
\begin{eqnarray}
\left| \frac{w_{m,n} t_{nR}}{E_m-\ep_R} \right| > 0.5
\end{eqnarray}
Likewise, the state $\ket{\psi_m}$ is determined to be degenerate with the left neighbor site when
\begin{eqnarray}
\left| \frac{w_{m,1} t_{L1}}{E_m-\ep_L} \right| > 0.5
\end{eqnarray}
Each cluster falls into one of the categories listed in Table \ref{table2}.

\begin{table*}
\begin{tabular}{cc|cc}
\multicolumn{2}{c}{number of degeneracies} & \multicolumn{2}{c}{renormalization procedure} \\
with $L$ & with $R$ & SDRG-RSnd & SDRG-RSdg \\ \hline
0 & 0 & \multicolumn{2}{c}{standard non-degenerate RS perturbation theory} \\
1 & 0 & no renormalization & left degeneracy \\ 
0 & 1 & no renormalization & right degeneracy \\
$> 1$ & 0 &  no renormalization & no renormalization \\
0 & $> 1$ &  no renormalization & no renormalization \\
$\ge 1$ & $\ge 1$ & no renormalization & no renormalization
\end{tabular}
\caption{\label{table2}Categories of degeneracies and how they are handled in the SDRG-RSnd and SDRG-RSdg code versions.
}
\end{table*}

The left and right degenerate renormalization procedures are described in the sections below.

\subsection{Equations for left degeneracy}

Let $\ket{\psi_{q_a}}$ be the state which is degenerate with $\ket{L}$.

$\ket{\psi_{q \ne q_a}}$ and $E_{q\ne q_a}$ as well as $\ket{R}$ and $E_R$ are renormalized using standard non-degenerate Rayleigh-Schr\"odinger perturbation theory as described in the body of the paper.

The following equations describe the renormalization of $\ket{\psi_{q_a}}$, $E_{q_a}$, $\ket{L}$, and $E_L$.
\begin{eqnarray}
\alpha &=& {y+ \sqrt{y^2 + h^2} \over \sqrt{h^2 + (y+ \sqrt{y^2+h^2})^2}} \\
\beta &=& {h \over \sqrt{h^2 + (y+ \sqrt{y^2+h^2})^2}} \\
{\rm where} \ x &=& {E_{q_a}+\ep_L \over 2} \\
y &=& {E_{q_a}-\ep_L \over 2} \\
h &=&w_{q_a,1} t_{L1}
\end{eqnarray}
\begin{eqnarray*}
\begin{array}{c|c}
E_{q_a}>\ep_L & E_{q_a}<\ep_L \\ \hline
y>0 & y<0 \\
\alpha > \beta & \alpha < \beta \\
\ket{\psi_+'} \rightarrow \ket{\psi_{q_a}'} & \ket{\psi_+'} \rightarrow \ket{L'} \\
\ket{\psi_-'} \rightarrow \ket{L'} & \ket{\psi_-'} \rightarrow \ket{\psi_{q_a}'}
\end{array}
\end{eqnarray*}
Energies 
\begin{eqnarray}
E_{+}'
&=& {E_{q_a} + \ep_L \over 2}
	+ \sqrt{\left( {E_{q_a} - \ep_L \over 2} \right)^2 + |w_{q_a,1} t_{L1}|^2} 
	\nonumber \\
	& &
	- { \left| \alpha w_{q_a,n} t_{nR} \right|^2
		\over \ep_R - E_+^{(0)} }
	- \sum_{q\ne q_a}
	{ \left| \beta w_{q,1}^* t_{L1} \right|^2
		\over E_q - E_+^{(0)} }
\label{EC_RGM_LDE+}
\\
E_{-}'
&=& {E_{q_a} + \ep_L \over 2}
	- \sqrt{\left( {E_{q_a} - \ep_L \over 2} \right)^2 + |w_{q_a,1} t_{L1}|^2}
	\nonumber \\
	& &
	- { \left| \beta w_{q_a,n} t_{nR} \right|^2 
		\over \ep_R - E_-^{(0)} }
	- \sum_{q\ne q_a}
	{ \left| \alpha w_{q,1}^* t_{L1} \right|^2 
		\over E_q - E_-^{(0)} }
\label{EC_RGM_LDE-}
\end{eqnarray}
States 
\begin{eqnarray}
\ket{\psi_+'} &=& \alpha \ket{\psi_{q_a}} + \beta \ket{L} 
	- {\alpha w_{q_a,n} t_{nR} \over \ep_R-E_{+}^{(0)} } \ket{R} 
	\nonumber \\
	& &
	- \sum_{q\ne q_a} { \beta w_{q,1}t_{L1}  \over E_q-E_{+}^{(0)} } \ket{\psi_q} 
\label{EC_RGM_LDS+}
\\
\ket{\psi_-'} &=& \beta \ket{\psi_{q_a}} - \alpha \ket{L} 
		- {\beta w_{q_a,n} t_{nR} \over \ep_R-E_{-}^{(0)} } \ket{R} 
	\nonumber \\
	& &
	+ \sum_{q\ne q_a} { \alpha w_{q,1}t_{L1}  \over E_q-E_{-}^{(0)} } \ket{\psi_q} 
\label{EC_RGM_LDS-}
\end{eqnarray}
We now address the hopping amplitude $t_{LR}$ and the IPR value of the degenerate cluster state.

Because the renormalized left state may be either the $\ket{\psi_+'}$ or $\ket{\psi_-'}$, both cases are considered.

When $E_{q_a}< \ep_L$, $y<0$, $\alpha<\beta$, 
	$\ket{\psi_+'} \rightarrow \ket{L'}$ and $\ket{\psi_-'} \rightarrow \ket{\psi_{q_a}'}$.
In this case, the hopping amplitude
\begin{eqnarray}
t_{LR}^+
&=& \bra{\psi_+'} H' \ket{R'} \\
&=& 
\alpha w_{q_a,n}^* t_{nR} 
	\left( 1 - {E_{q_a} \over E_{q_a}-\ep_R} - {\ep_R \over \ep_R - E_+^{(0)}} \right)
\nonumber \\ & &
- \beta t_{L1} t_{nR} 
	\left( \sum_{q=1}^n {w_{q,1} w_{q,n}^*  \over E_q-\ep_R} 
		+ \sum_{q\ne q_a} {w_{q,1} w_{q,n}^*  \over E_q - E_+^{(0)}} \right)
\nonumber \\ & & 
+ O(t/\Delta E)^2
\label{EC_RGM_LDtLR+}
\end{eqnarray}
At exact degeneracy the first term is zero.
As for the IPR, $\ket{\psi_{q_a}'}=\ket{\psi_-'}$, and using (\ref{EC_RGM_LDS-})
\begin{eqnarray}
IPR_{q_a,-} 
&=& 
\frac
{\sum_{i=1}^n \left( \beta w_{q_a,i} 
				+ \sum_{q\ne q_a} \frac{\alpha w_{q,1} t_{L1}}{E_q-E_-^{(0)}} w_{q,i} \right)^4
	+ \alpha^4
	+ \left( \frac{\beta w_{q_a,n}t_{nR}}{\ep_R-E_-^{(0)}} \right)^4}
{ \left( \sum_{i=1}^n \left( \beta w_{q_a,i} 
				+ \sum_{q\ne q_a} \frac{\alpha w_{q,1} t_{L1}}{E_q-E_-^{(0)}} w_{q,i} \right)^2
	+ \alpha^2
	+ \left( \frac{\beta w_{q_a,n}t_{nR}}{\ep_R-E_-^{(0)}} \right)^2 \right)^2 }
\label{EC_RGM_LD-IPR}
\end{eqnarray}

When $E_{q_a}> \ep_L$, $y>0$, $\alpha>\beta$, 
	$\ket{\psi_+'} \rightarrow \ket{\psi_{q_a}'}$ and $\ket{\psi_-'} \rightarrow \ket{L'}$.
In this case, the hopping amplitude
\begin{eqnarray}
t_{LR}^-
&=& \bra{\psi_-'} H' \ket{R'} \\
&=& \beta w_{q_a,n} t_{nR} \left( 1 - {E_{q_a} \over E_{q_a} - \ep_R} - {\ep_R \over \ep_R - E_-^{(0)}} \right) 
\nonumber \\ & & 
+ \alpha t_{L1} t_{nR} \left( \sum_{q=1}^n {w_{q,1} w_{q,n}^* \over E_q - \ep_R}
					+ \sum_{q \ne q_a} {w_{q,1} w_{q,n}^* \over E_q - E_-^{(0)}} \right)
\nonumber \\ & & 
+ O(t/\Delta E)^2					
\label{EC_RGM_LDtLR-}
\end{eqnarray}
Again, at exact degeneracy the first term is zero.
As for the IPR, $\ket{\psi_{q_a}'}=\ket{\psi_+'}$, and using (\ref{EC_RGM_LDS+})
\begin{eqnarray}
IPR_{q_a,+} 
&=& 
\frac
{\sum_{i=1}^n \left( \alpha w_{q_a,i} 
				- \sum_{q\ne q_a} \frac{\beta w_{q,1} t_{L1}}{E_q-E_+^{(0)}} w_{q,i} \right)^4
	+ \beta^4
	+ \left( \frac{\alpha w_{q_a,n}t_{nR}}{\ep_R-E_+^{(0)}} \right)^4}
{ \left( \sum_{i=1}^n \left( \alpha w_{q_a,i} 
				- \sum_{q\ne q_a} \frac{\beta w_{q,1} t_{L1}}{E_q-E_+^{(0)}} w_{q,i} \right)^2
	+ \beta^2
	+ \left( \frac{\alpha w_{q_a,n}t_{nR}}{\ep_R-E_+^{(0)}} \right)^2 \right)^2 }
\label{EC_RGM_LD+IPR}
\end{eqnarray}

\subsection{Equations for right degeneracy}

Let $\ket{\psi_{q_a}}$ be the state which is degenerate with $\ket{R}$.

$\ket{\psi_{q \ne q_a}}$ and $E_{q\ne q_a}$ s well as $\ket{L}$ and $E_L$ are renormalized using standard non-degenerate Rayleigh-Schr\"odinger perturbation theory as described in the body of the paper.

The following equations describe the renormalization of $\ket{\psi_{q_a}}$, $E_{q_a}$, $\ket{R}$, and $E_R$.
\begin{eqnarray}
\alpha &=& {y+ \sqrt{y^2 + h^2} \over \sqrt{h^2 + (y+ \sqrt{y^2+h^2})^2}} \\
\beta &=& {h \over \sqrt{h^2 + (y+ \sqrt{y^2+h^2})^2}} \\
{\rm where} \ x &=& {E_{q_a}+\ep_R \over 2} \\
y &=& {E_{q_a}-\ep_R \over 2} \\
h &=& w_{q_a,n} t_{nR}
\end{eqnarray}
\begin{eqnarray*}
\begin{array}{c|c}
E_{q_a}>\ep_R & E_{q_a}<\ep_R \\ \hline
y>0 & y<0 \\
\alpha > \beta & \alpha < \beta \\
\ket{\psi_+'} \rightarrow \ket{\psi_{q_a}'} & \ket{\psi_+'} \rightarrow \ket{R'} \\
\ket{\psi_-'} \rightarrow \ket{R'} & \ket{\psi_-'} \rightarrow \ket{\psi_{q_a}'}
\end{array}
\end{eqnarray*}
Energies
\begin{eqnarray}
E_+' &=& {E_{q_a} + \ep_R \over 2}
		+ \sqrt{\left( {E_{q_a} - \ep_R \over 2} \right)^2 + |w_{q_a,n} t_{nR}|^2}
\nonumber \\ & &
		- { \left| \alpha w_{q_a,1} t_{L1} \right|^2
			\over \ep_L - E_+^{(0)} }
		- \sum_{q\ne q_a}
		{ \left| \beta w_{q,n}^* t_{nR} \right|^2
			\over E_q - E_+^{(0)} }
\label{EC_RGM_RDE+}
\\
E_-' &=& {E_{q_a} + \ep_R \over 2}
		- \sqrt{\left( {E_{q_a} - \ep_R \over 2} \right)^2 + |w_{q_a,n} t_{nR}|^2}
\nonumber \\ & &
		- { \left| \beta w_{q_a,1} t_{L1} \right|^2 
			\over \ep_L - E_-^{(0)} }
		- \sum_{q\ne q_a}
		{ \left| \alpha w_{q,n}^* t_{nR} \right|^2 
			\over E_q - E_-^{(0)} }
\label{EC_RGM_RDE-}
\end{eqnarray}
States
\begin{eqnarray}
\ket{\psi_+'} &=& \alpha \ket{\psi_{q_a}} + \beta \ket{R}
			- {\alpha w_{q_a,1} t_{L1} \over \ep_L-E_{+}^{(0)} } \ket{L} 
\nonumber \\ & &
			- \sum_{q\ne q_a} { \beta w_{q,n}^* t_{nR}  
							\over E_q-E_{+}^{(0)} } \ket{\psi_q} 
\label{EC_RGM_RDS+}
\\
\ket{\psi_-'} &=& \beta \ket{\psi_{q_a}} - \alpha \ket{R}
			- {\beta w_{q_a,1} t_{L1} \over \ep_L-E_{-}^{(0)} } \ket{L} 
\nonumber \\ & &
			+ \sum_{q\ne q_a} { \alpha w_{q,n}^*t_{nR}  
							\over E_q-E_{-}^{(0)} \ket{\psi_q}} 
\label{EC_RGM_RDS-}
\end{eqnarray}
We now address the hopping amplitude $t_{LR}$ and the IPR value of the degenerate cluster state.

Because the renormalized right state may be either the $\ket{\psi_+'}$ or $\ket{\psi_-'}$, both cases are considered.

When $E_{q_a}<\ep_R$, $y<0$, $\alpha<\beta$, $\ket{\psi_+'} \rightarrow \ket{R'}$ and $\ket{\psi_-'} \rightarrow \ket{\psi_{q_a}'}$.  In this case, the hopping amplitude
\begin{eqnarray}
t_{LR}^+ &=& \bra{L'} H' \ket{\psi_+'} \\
&=& \alpha w_{q_a,1} t_{L1} 
	\left( 1 - {\ep_L \over \ep_L - E_+^{(0)}} - {E_{q_a} \over E_{q_a} - \ep_L} \right)
\nonumber \\ & & 
- \beta t_{L1} t_{nR} 
	\left( \sum_{q\ne q_a} {w_{q,1} w_{q,n}^* \over E_q - E_+^{(0)}}
		+ \sum_{q=1}^n {w_{q,1} w_{q,n}^* \over E_q - \ep_L} \right)
\nonumber \\ & &
+ O(t/\Delta E)^2
\end{eqnarray}
At exact degeneracy the first term is zero.
As for the IPR, $\ket{\psi_{q_a}'}=\ket{\psi_-'}$, and using (\ref{EC_RGM_RDS-})
\begin{eqnarray}
IPR_{q_a,-} 
&=& 
\frac
{\sum_{i=1}^n \left( \beta w_{q_a,i} 
				+ \sum_{q\ne q_a} \frac{\alpha w_{q,n}^* t_{nR}}{E_q-E_-^{(0)}} w_{q,i} \right)^4
	+ \alpha^4
	+ \left( \frac{\beta w_{q_a,1}t_{L1}}{\ep_L-E_-^{(0)}} \right)^4}
{ \left( \sum_{i=1}^n \left( \beta w_{q_a,i} 
				+ \sum_{q\ne q_a} \frac{\alpha w_{q,n}^* t_{nR}}{E_q-E_-^{(0)}} w_{q,i} \right)^2
	+ \alpha^2
	+ \left( \frac{\beta w_{q_a,1}t_{L1}}{\ep_L-E_-^{(0)}} \right)^2 \right)^2 }
\label{EC_RGM_RD-IPR}
\end{eqnarray}

When $E_{q_a}>\ep_R$, $y>0$, $\alpha>\beta$, $\ket{\psi_-'} \rightarrow \ket{R'}$ and $\ket{\psi_+'} \rightarrow \ket{\psi_{q_a}'}$.  In this case, the hopping amplitude
\begin{eqnarray}
t_{LR}^- &=& \bra{L'} H' \ket{\psi_-'} \\
&=& 
\beta w_{q_a,1} t_{L1} 
	\left( 1 - {\ep_L \over \ep_L - E_-^{(0)}} - {E_{q_a} \over E_{q_a} - \ep_L} \right)
\nonumber \\ & &
+ \alpha t_{L1} t_{nR} 
	\left( \sum_{q \ne q_a} {w_{q,1} w_{q,n}^* \over E_q - E_-^{(0)}}
		+ \sum_{q=1}^n {w_{q,1} w_{q,n}^* \over E_q - \ep_L} \right)
\nonumber \\ & &
+ O(t/\Delta E)^2
\label{EC_RGM_RDtLR-}
\end{eqnarray}
Again, at exact degeneracy the first term here is zero.
As for the IPR, $\ket{\psi_{q_a}'}=\ket{\psi_+'}$, and using (\ref{EC_RGM_RDS+})
\begin{eqnarray}
IPR_{q_a,+} 
&=& 
\frac
{\sum_{i=1}^n \left( \alpha w_{q_a,i} 
				- \sum_{q\ne q_a} \frac{\beta w_{q,n}^* t_{nR}}{E_q-E_+^{(0)}} w_{q,i} \right)^4
	+ \beta^4
	+ \left( \frac{\alpha w_{q_a,1}t_{L1}}{\ep_L-E_+^{(0)}} \right)^4}
{ \left( \sum_{i=1}^n \left( \alpha w_{q_a,i} 
				- \sum_{q\ne q_a} \frac{\beta w_{q,n}^* t_{nR}}{E_q-E_+^{(0)}} w_{q,i} \right)^2
	+ \beta^2
	+ \left( \frac{\alpha w_{q_a,1}t_{L1}}{\ep_L-E_+^{(0)}} \right)^2 \right)^2 }
\label{EC_RGM_RD+IPR}
\end{eqnarray}

\section{Orthogonality of the basis}
\label{app:basis}

We begin with a basis of states $\ket{i}$ for $i=1$ to $N$. 
These states or orthonormal:
\begin{eqnarray}
\bra{i}\ket{j} &=& \delta_{ij}
\end{eqnarray}
When we renormalize a cluster $i=C1$ to $Cn$, we record its energies in the DOS and its states in the IPR, after which the states $\ket{C1}$ to $\ket{Cn}$ no longer appear explicitly in our calculations.
This is referred to as ``removing" the cluster.
We are left with a modified basis of states which includes
\begin{eqnarray}
\ket{1}, ..., \ket{L-1}, \ket{L'}, \ket{R'}, \ket{R+1}, ... , \ket{N}
\end{eqnarray}
Is this basis orthonormal?
Only the states $\ket{L'}$ and $\ket{R'}$ have changed in the renormalization process, so for all the other states if they were orthonormal before then they remain so.
Consider therefore just the inner products involving $\ket{L'}$ and $\ket{R'}$.  

Single-site cluster
\begin{eqnarray}
\bra{L'} \ket{L'} &=& \left( \bra{L} - {t_{L1} \over \ep_1-\ep_L} \bra{1} \right)
				\left( \ket{L} - {t_{L1} \over \ep_1 - \ep_L} \ket{1} \right) \\
	&=& 1 + O(t/\Delta E)^2
\\
\bra{L'} \ket{R'} &=& \left( \bra{L} - {t_{L1} \over \ep_1-\ep_L} \bra{1} \right)
			\left( \ket{R} - {t_{1R} \over \ep_1 - \ep_R} \ket{1} \right) \\
	&=& 0 + O(t/\Delta E)^2 \\
\bra{L'}\ket{j} &=& 0 \ \forall \ j \ne L',R' 
\end{eqnarray}

Multi-site cluster
\begin{eqnarray}
\bra{L'} \ket{L'} &=& \left( \bra{L} - \sum_q {w_{q,1} t_{L1} \over E_q-\ep_L} \bra{\psi_q} \right) \nonumber \\
			& & \times
			\left( \ket{L} - \sum_q {w_{q,1}^* t_{L1} \over E_q-\ep_L} \ket{\psi_q} \right) \\
	&=& 1 + O(t/\Delta E)^2
\\
\bra{L'} \ket{R'} &=& \left( \bra{L} - \sum_q {w_{q,1} t_{L1} \over E_q-\ep_L} \bra{\psi_q} \right) \nonumber \\
			& & \times
			\left( \ket{R} - \sum_q {w_{q,n}^* t_{nR} \over E_q-\ep_R} \ket{\psi_q} \right) \\
	&=& 0 + O(t/\Delta E)^2
\end{eqnarray}
Similar results obtain for $\ket{R'}$.
Therefore, to first order in $t/\Delta E$, there is no loss of orthonormality in the basis.  

\section{Comparison with results shown in \jbn}
\label{app:compare}

\begin{figure}[htbp] 
\includegraphics[width=\columnwidth]{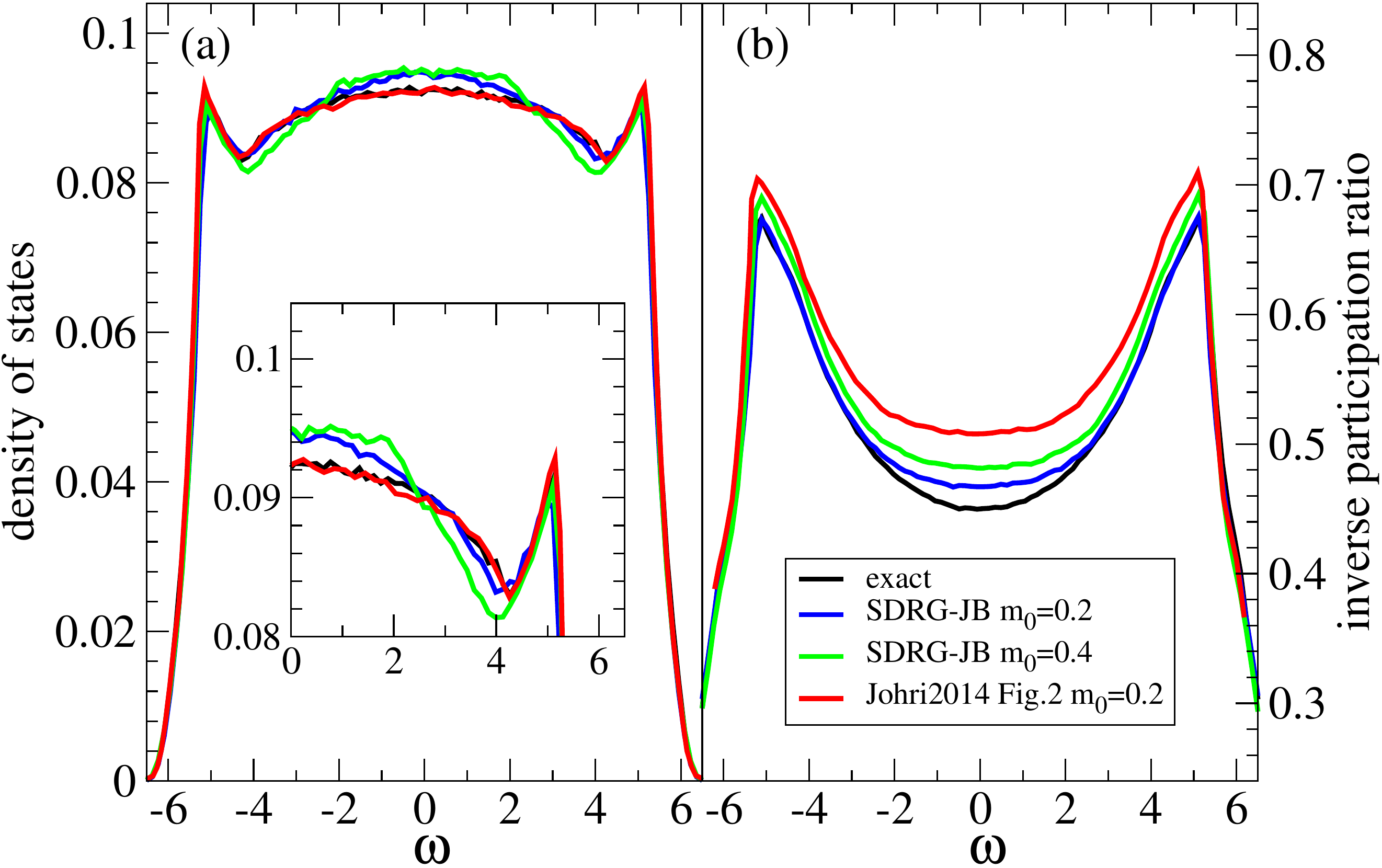} 
\caption{\label{dosipr_compare}Density of states and inverse participation ratio at disorder strength $W=10$, comparing the results of our implementation of the procedure proposed in \jb and the results presented in \jbn.  
Exact and SDRG-JB results are for $N_{ss}=10^4$ and $N_{dc}=10^3$.
The results from \jb have been extracted from Fig.\ 2 of their paper using the digitizing software WebPlotDigitizer.
Specifically, these are the data labeled $m_0=0.2$ and represented by stars.
}
\end{figure}

Fig.\ \ref{dosipr_compare} provides a direct comparison between the results of our implementation of the method described in \jb and the results presented in \jbn.
One conclusion from this figure is that our calculation and that of \jb are not the same.
Our implementation is described in detail in the body of the paper with key equations summarized in Table \ref{table1}.
Potential points are difference are (i) we keep only terms first order in $t/\Delta E$, and (ii) the form of $t_{LR}$ in the case of multi-site clusters.
\jb does not provide an expression for $t_{LR}$ in this case.
We first tried strictly following the equations in \jb, using $t_{LR}=\bra{L'}H\ket{R'}$.  However, this is zero for clusters of more than one site, to first order in $t/\Delta E$, and \jb Fig.\ 3 shows the generation of nonzero bonds for all cluster sizes.
We therefore constructed an alternative expression, shown in Table \ref{table1}, which (i) follows the form of other \jb equations, (ii) reduces to the one-site cluster expression in the limit $n\to 1$,  and (iii) produces similar distributions of bond strengths to those shown in \jb Fig.\ 3.  

Also apparent in this comparison between the results of our implementation SDRG-JB and the results presented in \jb is that the \jb results are closer to the exact results for the DOS but the SDRG-JB results are closer to the exact results for the IPR.  
Meanwhile, as shown in the body of the paper, the procedures based on standard Rayleigh-Schr\"odinger perturbation theory (SDRG-RSnd and SDRG-RSdg) produce DOS results much closer to the exact results than SDRG-JB,
and the SDRG-RSdg inverse participation results are slightly better than those from SDRG-JB.


\begin{thebibliography}{10}
\expandafter\ifx\csname url\endcsname\relax
  \def\url#1{\texttt{#1}}\fi
\expandafter\ifx\csname urlprefix\endcsname\relax\def\urlprefix{URL }\fi
\expandafter\ifx\csname href\endcsname\relax
  \def\href#1#2{#2} \def\path#1{#1}\fi

\bibitem{Anderson1958}
P.~Anderson, Absence of diffusion in certain random lattices, Physical Review
  109 (1958) 1492.

\bibitem{Abrahams1979}
E.~Abrahams, P.~Anderson, D.~Licciardello, T.~Ramakrishnan, Scaling theory of
  localization: Absence of quantum diffusion in two dimensions, Physical Review
  Letters 42 (1979) 673.

\bibitem{Gopalakrishnan2020}
S.~Gopalakrishnan, S.~Parameswaran, Dynamics and transport at the threshold of
  many-body localization, Physics Reports 862 (2020) 1.

\bibitem{Abanin2019}
D.~Abanin, E.~Altman, I.~Bloch, M.~Serbyn, Colloquium: Many-body localization,
  thermalization, and entanglement, Reviews of Modern Physics 91 (2019) 021001.

\bibitem{Nandkishore2015}
R.~Nandkishore, D.~Huse, Many-body localization and thermalization in quantum
  statistical mechanics, Annual Review of Condensed Matter Physics 6 (2015) 15.

\bibitem{Ma1979}
S.~Ma, C.~Dasgupta, C.~Hu, Random antiferromagnetic chain, Physical Review
  Letters 43 (1979) 1434.

\bibitem{Dasgupta1980}
C.~Dasgupta, S.~Ma, Low-temperature properties of the random heisenberg
  antiferromagnetic chain, Physical Review B 22 (1980) 1305.

\bibitem{Fisher1994}
D.~Fisher, Random antiferromagnetic quantum spin chains, Physical Review B 50
  (1994) 3799.

\bibitem{Igloi2005}
F.~Igloi, C.~Monthus, Strong disorder rg approach of random systems, Physics
  Reports 412 (2005) 277.

\bibitem{Igloi2018}
F.~Igloi, C.~Monthus, Strong disorder rg approach --- a short review of recent
  developments, European Physical Journal B 91 (2018) 290.

\bibitem{Wilson1974}
K.~Wilson, J.~Kogut, The renormalization group and the epsilon expansion,
  Physics Reports 12 (1974) 75.

\bibitem{Monthus2009}
C.~Monthus, T.~Garel, Statistics of renormalized on-site energies and
  renormalized hoppings for anderson localization in two and three dimensions,
  Physical Review B 80~(024203) (2009).

\bibitem{Mard2014}
H.~J. Mard, J.~A. Hoyos, E.~Miranda, V.~Dobrosavljevic, Strong-disorder
  renormalization-group study of the one-dimensional tight-binding model,
  Physical Review B 90~(125141) (2014).

\bibitem{Mard2017}
H.~J. Mard, J.~A. Hoyos, E.~Miranda, V.~Dobrosavljevic, Strong-disorder
  approach for the anderson localization transition, Physical Review B
  96~(045143) (2017).

\bibitem{Johri2014}
S.~Johri, R.~Bhatt, Large-disorder renormalization group study of the anderson
  model of localization, Physical Review B 90 (2014) 060205(R).

\bibitem{Johri2012a}
S.~Johri, R.~Bhatt, Singular behavior of eigenstates in anderson's model of
  localization, Physical Review Letters 109 (2012) 076402.

\bibitem{Johri2012b}
S.~Johri, R.~Bhatt, Singular behavior of anderson-localized wave functions for
  a two-site model, Physical Review B 86 (2012) 125140.

\end{thebibliography}

\end{document}